\newcommand{\Om}{\Omega}			

\newcommand{\OmGVM}{{\Omega_\mathrm{gvm}}}

\newcommand{\OmGVS}{{\Omega_\mathrm{gvs}}}
\newcommand{\OmGVD}{{\Omega_\mathrm{gvd}}}

\newcommand{\dtA}{{\tau_\mathrm{gvm}}}

\newcommand{\tauGVM}{{\tau_\mathrm{gvm}}}
\newcommand{\tauGVS}{{\tau_\mathrm{gvs}}}
\newcommand{\tauGVD}{{\tau_\mathrm{gvd}}}

\newcommand{\rrOm}{ {  r} (\Omega) } 
\newcommand{\rr} { { r} }

\newcommand{\A} {\hat{A} }

\newcommand{\sinc}{{\rm sinc}}

\newcommand{\nn}{\nonumber}
\newcommand{\bsub}{\begin{subequations}}
\newcommand{\esub}{\end{subequations}}
\newcommand{\beq}{\begin{equation}}
\newcommand{\eeq}{\end{equation}}
\newcommand{\beqa}{\begin{eqnarray}}
\newcommand{\eeqa}{\end{eqnarray}}
\newcommand{\beql}{\begin{subequations}\begin{eqnarray}}
\newcommand{\eeql}{\end{eqnarray}\end{subequations}}
\documentclass[pra,aps,amsmath, nofootinbib]{revtex4}
\usepackage{amsmath} 
\usepackage{graphicx}
\usepackage{braket}
\usepackage{color}
\usepackage{cases}
\begin{document}

\markboth{A.Gatti and E-Brambilla}
{ CV entanglement in the  MOPO}
\title{Continuous variable entanglement of counter-propagating twin beams}
\author{ A.~Gatti$^{1,2}$ and  E.~Brambilla$^{2}$ }
\address{$1$ Istituto di Fotonica e Nanotecnologie del CNR, Piazza Leonardo da Vinci 32, Milano, Italy \\ $^2$ Dipartimento di Scienza e Alta Tecnologia dell'Universit\`a dell'Insubria, Via Valleggio 11 Como, Italy}
\begin{abstract}
This work  describes the continuous-variable entanglement of the counter-propagating  twin beams generated in a Mirrorless Optical Parametric Oscillator below threshold, encompassing both their  quadrature and photon-number correlation.  In the first case,   a comparison with the single-pass co-propagating geometry outlines 
a completely different stability of the two sources with respect to the phase-angle.  In the second case, stimulated by the  critical divergence of the correlation time evidenced by {\em Corti et al. }, we address the issue of the temporal bandwidth of the intensity squeezing. 
\end{abstract}
\maketitle
\section{Introduction}
\label{sec:intro}
\begin{figure}[ht]
\centering 
     \includegraphics[width=0.7\textwidth]{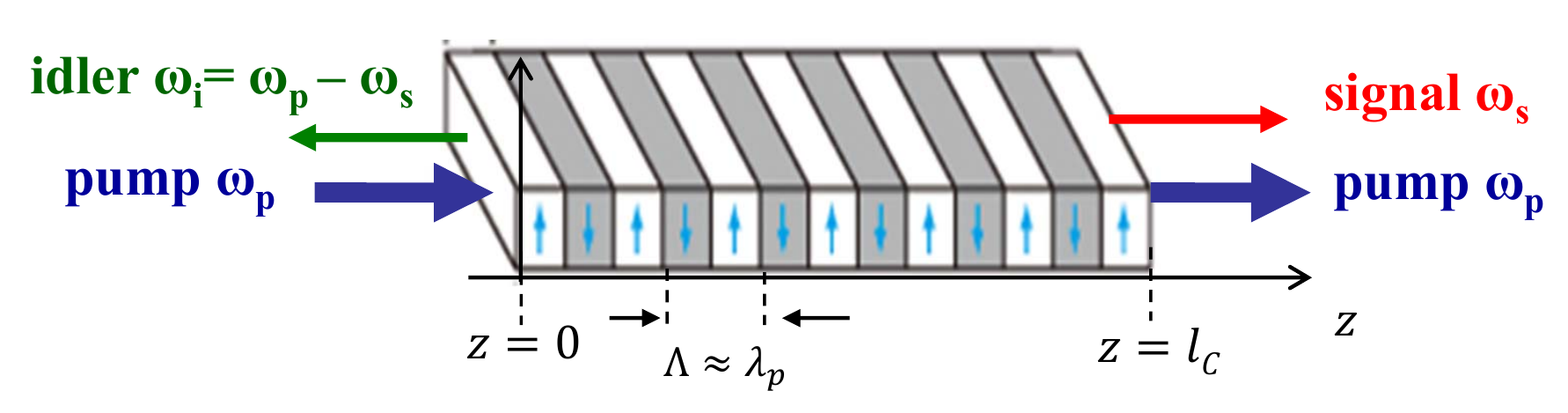}
           \caption{(Color online) Scheme of counterpropagating parametric down-conversion, taking place in a periodically poled crystal of length $l_c$ and poling period $\Lambda\approx \lambda_p$. Quasi-phase matching  requires that the idler field is generated in the backward direction with respect to the signal and pump.}
\label{fig_scheme}
\end{figure}

Squeezed light and continuous variable entanglement are precious resources for quantum  information, communication  and metrology.  One of the most accessible and widely used source is represented by  the twin beams \cite{Rarity1992} generated through parametric-down conversion (PDC) from a pump laser. 
This work focuses on a peculiar  configuration, where the twin beams are generated in opposite directions, and counter-propagate in a slab of   $\chi^{(2)}$  material (Fig.\ref{fig_scheme}). This process is allowed only in the presence of quasi-phase matching in periodically poled  crystals, and presents the challenge of requiring very short poling periods  \cite{Busacca2002,Canalias2003} on the order of the pump wavelength. Predicted in the sixties\cite{Harris1966},  counter-propagating PDC had indeed to wait almost forty years before being demonstrated \cite{Canalias2007}. 
\par
Counter-propagating PDC emerged in the last years as a  promising source of quantum light, with  several peculiar and appealing features\cite{Christ2009,Gatti2015,Gatti2017b, Corti2016, Gatti2017}.
First of all, in contrast with the usual single-pass co-propagating geometry, this source is  narrowband, so that in the spontaneous regime it has  the potentiality to 
 generate  narrowband  heralded single photons in almost pure states\cite{Christ2009,Gatti2015, Gatti2017b}.
The second peculiarity is the presence of a threshold value of the pump intensity, beyond which the system makes a transition to coherent oscillations, similarly to what happens in a standard Optical Parametric Oscillator (OPO), from which the name  {\em Mirrorless Optical parametric Oscillator}(MOPO) \cite{Canalias2007}. Responsible of this critical behaviour is a feedback mechanism, which in this case is  established not by the cavity mirrors, but by  the back-propagating wave, in combination  with stimulated down-conversion\cite{Corti2016}.   A recent analysis\cite{Gatti2017} has shown that this cavityless configuration  of PDC may produce the same high and stable  level of squeezing and quadrature correlation as the  OPO, and may thus become a robust and monolithic alternative to the cavity configuration.   
\par
This work  provides a general description of the continuous-variable entanglement of the MOPO twin beams below threshold, encompassing both their  quadrature and photon-number correlation.  In the first case, 
 a comparison will be performed with the single-pass co-propagating geometry, which in its high-gain regime can be used as a source of squeezed light \cite{Chekhova2015, Eto2008,Kaiser2016}. A part from the huge difference of the bandwidth involved, our analysis will outline a completely different stability of the two sources with respect to the phase-angle at which squeezing takes place. 
For the photon number correlation, our analysis will address the question  of how long the twin beams should be detected in order to observe sub-shot noise fluctuations in the difference of the their photon-numbers, or in other words, the issue  of the bandwidth of the intensity squeezing. The interest in this sense is stimulated by the findings of Ref.\cite{Corti2016}, where it was shown that on approaching the MOPO threshold the twin beams become correlated over a longer and longer time, ideally diverging at threshold. 
\par
The work is organized as follow: after briefly introducing the quantum model for the device, and describing the spectral characteristics of the emission (sections \ref{sec:model} and  \ref{sec:bandwidth}), Sec.\ref{sec:state}  reviews some general properties which are common to all processes of photon-pair generation. This  allows in the next section \ref{sec:quad} a straightforward analysis  of the quadrature correlation in the two counter-propagating and co-propagating configurations. The final Sec \ref{sec:intensity}  addresses the problem of the intensity squeezing in the MOPO and of its bandwidth.

\section{The model}
\label{sec:model}
\begin{figure}[ht]
\centering
     \includegraphics[width=0.6\textwidth]{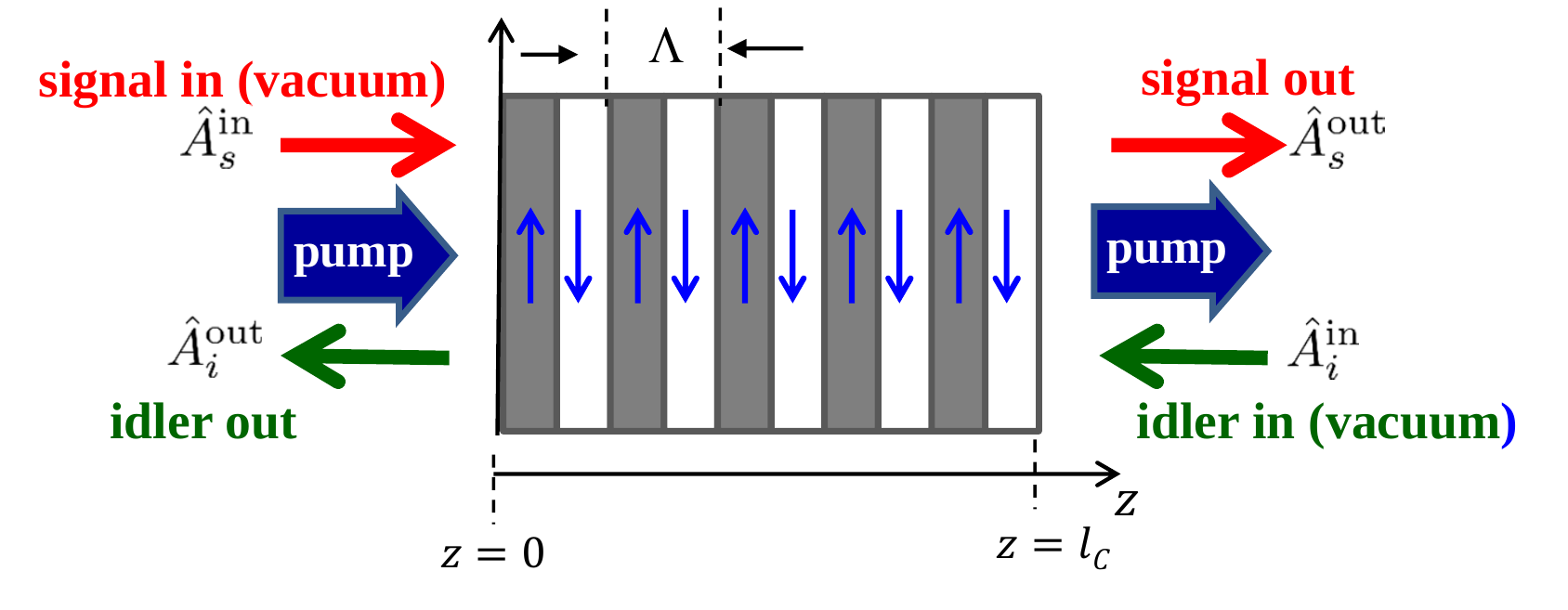}
           \caption{(Color online) Counterpropagating input-otput scheme.}
\label{fig2}
\end{figure}

We consider the  geometry  in Fig.\ref{fig_scheme},  in which the laser pump and the down-converted signal  co-propagate along the $+z$ direction, while the idler back-propagates in the $-z$ direction along a periodically poled slab of a $\chi^{(2)}$ material.  Our quantum model for this configuration is described  in Refs.\cite{Corti2016,Gatti2015,Gatti2017} 
(see also \cite{Suhara2010}, \cite{Ding1996}).
We summarize in the following the main points \\
$-$  A purely temporal description of the twin beams is carried out, assuming  either a  waveguided configuration or that a small angular bandwidth is collected.\\
$-$ The pump laser is described as a monochromatic classical beam of frequency $\omega_p$  and amplitude   $\alpha_p=
\sqrt{I_p} e^{i\phi_p} $. Below the MOPO threshold, we assume that it is  undepleted by the parametric interaction, an   approximation which clearly becomes unphysical as one gets very close to the MOPO threshold. \\
$-$  The signal and idler fields are  described by   quantum field operators  for two wavepackets centered around 
 frequencies $\omega_s$ and $\omega_i=\omega_p-\omega_s$, such that their corresponding wave numbers in the medium,
$k_j={\omega_j \, n_j(\omega_j)}/{c}$,   satisfy the quasi-phase matching condition 
$
k_s-k_i = k_p-k_G
$
where $k_G=2\pi m /\Lambda$ is the reciprocal vector of the nonlinear grating at first or low order, $m=1,3,5$.
\\
$-$
The generation of twin beams in the nonlinear slab is then described by linear parametric equations, that couple the  signal and idler field operators  via the dimensionless gain parameter 
\beq
g= \sqrt{2\pi} \chi |\alpha_p| l_c ,
\label{gain}
\eeq
where $\chi$ is proportional to the $\chi^{(2)}$ nonlinear susceptibility of the medium and $l_c$ is the crystal length. \\
These propagation equations are then solved   in terms of linear input-output transformations linking the output field  operators to the input ones. Notice that here the boundaries differ from the standard ones, because the output signal and idler fields: $\hat{A}_s^{\text{out}}(\Omega) = \hat A_s (\Omega, z=l_c) $, 
$ \hat{A}_i^{\text{out}}(\Omega) =  \hat A_i (\Omega, z=0) $,  appear on the opposite faces of the slab (Fig.\ref{fig2}), while the input fields: $\hat{A}_s^{\text{in}}(\Omega)=  \hat{A}_s(\Omega,z=0 ) $, 
$\hat{A}_i^{\text{in}}(\Omega)= \hat{A}_i(\Omega,z=l_c )$, assumed in the vacuum state, enter from opposite faces. 
The input-otput relations have the general form of a  Bogoliubov transformation, characteristic of processes where particles are generated in pairs:
\bsub
\begin{align}
\hat{A}_s^{\text{out}}(\Omega)&=U_s(\Omega)\hat{A}_s^{\text{in}}(\Omega)+V_s(\Omega)\hat{A}_i^{\text{in} \dagger}(-\Omega)\\
\hat{A}_i^{\text{out}}(-\Omega)&=U_i(-\Omega)\hat{A}_i^{\text{in}}(-\Omega)+V_i(-\Omega)\hat{A}_s^{\text{in}\dagger}(\Omega).
\end{align}
\label{inout}
\esub
Here $\hat{A}_j^{\text{out,in}}(\Omega) $  $j=,s,i$ are the positive frequency parts of the electric field operators, with dimensions of photon annihilation operators, such that $  \hat{A}_j^{\dagger \text{out}}(\Omega)  \hat{A}_j^{\text{out}}(\Omega) $ are the output photon numbers per unit frequency. 
 Capital $\Omega$ denotes the frequency offset from the respective central frequencies $\omega_s$ and $\omega_i$,  

Unlike the co-propagating case, where the coefficients of the input-output transformation grow exponentially with the propagation length in the medium, in the MOPO case the  coefficients $U_j(\Omega)$ and $V_j (\Omega)$ are   trigonometric functions of the crystal length (through the gain parameter $g \propto l_c$),  and read\cite{Corti2016, Suhara2010}:
\bsub
\begin{align}
&U_s(\Omega)=e^{i k_s l_c}e^{i\beta(\Omega)}\phi(\Omega) ;   & &
V_s(\Omega)=e^{i (k_s-k_i) l_c}g e^{i\phi_p}\frac{\sin\gamma(\Omega)}{\gamma(\Omega)}\phi(\Omega); \\
&U_i(-\Omega)=e^{i k_i l_c}e^{i\beta(\Omega)}\phi^*(\Omega);  & & 
V_i(-\Omega)= ge^{i\phi_p}\frac{\sin\gamma(\Omega)}{\gamma(\Omega)}\phi^*(\Omega).
\end{align}
\label{UV}
\esub
where:
\begin{align}
\phi(\Omega)&=\frac{1}{\cos\gamma(\Omega)-i\frac{\bar{\mathcal{D}}(\Omega)l_c}{2\gamma(\Omega)}\sin\gamma(\Omega)}\\
\gamma(\Omega)&=\sqrt{g^2+\frac{\bar{\mathcal{D}}^2(\Omega)l_c^2}{4}},
\label{gamma} \\
{\mathcal{D}}(\Omega)&=k_{s}(\Omega)-k_{i}(-\Omega)-k_p+k_G, 
\label{DD}\\
\beta(\Omega)&=[k_s(\Omega)+k_i(-\Omega)-(k_s+k_i)]\frac{l_c}{2}\label{beta}
\end{align}
The function $ {\mathcal{D}}(\Om)$ in Eq.\eqref{DD} is the phase mismatch of  the two frequency conjugate waves at $\omega_s +\Om$ and $\omega_i -\Om$,  with 
$k_j (\Om) = n_j (\Om) (\omega_j + \Om) /c$  being  their wave numbers in the medium.  The function $\beta (\Om)$  in Eq.\eqref{beta} is instead a global propagation phase. 
\\
As can be easily checked, these coefficients  satisfy the unitarity conditions
\bsub
\begin{align}
&|U_j(\Omega)|^2-|V_j(\Omega)|^2=1,\;\;\;\;\;\;j=s,i\\
&U_s(\Omega)V_i(-\Omega)=U_i(-\Omega)V_s(\Omega)
\end{align}
\label{unitarity}
\esub
Most importantly,   $U_j(\Omega)$ and $V_j(\Omega)$ diverge when approaching 
\beq
g=g_{\mathrm{thr}} = \frac{\pi}{2}\; , 
\eeq
the value of the parametric gain corresponding to the MOPO threshold in the CW pump regime \cite{Ding1996}.

\section{Characteristic spectral bandwidths}
\label{sec:bandwidth}
\begin{figure}[ht]
     \includegraphics[width=0.9\textwidth]{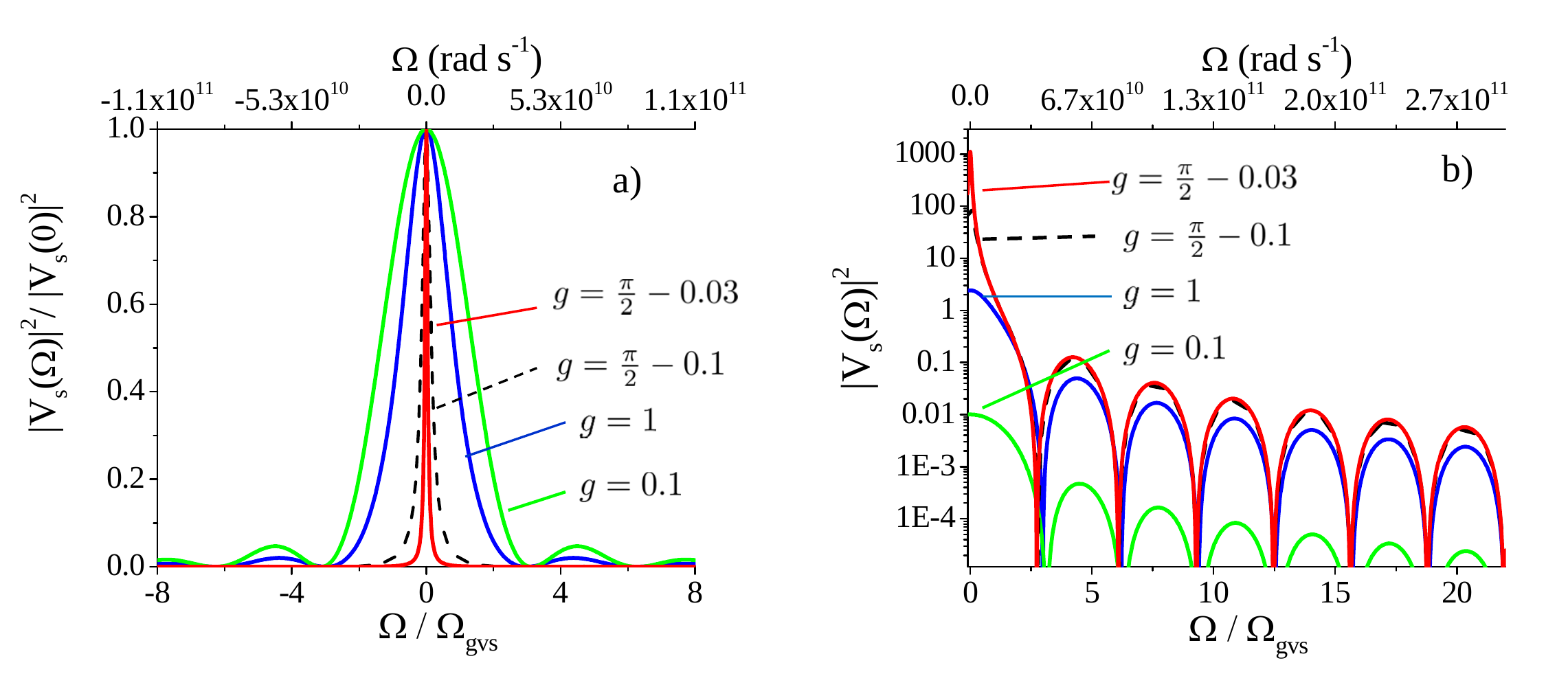}
           \caption{(Color online) Intensity spectra $|V_{s}(\Omega) |^2    $  \eqref{Vs}, plotted for different values of the gain $g$, as a function of  $\Omega/\OmGVS$ (bottom axis) and of the frequency in a PPLN slab  poled for the counterpropagating type 0 interaction. In  a) the spectra are normalized to their peak value, in  b)  they are shown in logarithmic scale. $\lambda_p =771$nm,   $\lambda_s= \lambda_i = 1542$nm, $\Lambda = 212.7$nm, $l_c=1cm$.}
\label{fig3_spectra}
\end{figure}

A Taylor expansion of the  phase mismatch ${\mathcal{D}}(\Omega)$ (\ref{DD}) in series of the frequency offset $\Om$ gives: 
\begin{align}
\frac{ {\mathcal{D}}(\Omega)l_c}{2}&=\frac{l_c}{2}(k_s'+k_i')\Omega+\frac{l_c}{4}(k_s''-k_i'')\Omega^2+\cdots
\label{Taylor_D}\\
&\simeq \frac{l_c}{2}(k_s'+k_i')\Omega: = \tauGVS \Omega
\label{linear}
\end{align}
where $k'_j, k_j''$ indicate  derivatives of the wavenumbers $k_j(\Om)$, calculated at the reference frequencies $\Om=0$, and 
\beq
  \tauGVS= \frac{1}{2}\left[\frac{l_c}{v_{gs}}+\frac{l_c}{v_{gi}}\right] \equiv  \OmGVS^{-1}
\eeq
is a time scale characteristic of counterpropagating interactions, 
 involving the { sum} of the inverse group velocities $v_{gj }= 1/k'_j$ .  This  {\em long} time scale roughly corresponds to the maximal delay that may occur between the exits  of  two twins down-converted from the same pump photon, and  is on the order of the  transit time of light along  the slab because they appear at  its opposite sides   \cite{Corti2016, Gatti2015}.  Its inverse  $\OmGVS$ is responsible  the narrow width of the spectrum of downconverted light,  both below  \cite{Suhara2010, Christ2009, Gatti2015, Corti2016}, and above the MOPO threshold 
 \cite{Canalias2007}.  The first term in the Taylor expansion \eqref{Taylor_D} is by far the dominant one, so that the  linear approximation for the phase mismatch in Eq. \eqref{linear}  is well justified. 
\\
Figure \ref{fig3_spectra} shows an example of such spectra, calculated for various values of $g$ below the MOPO threshold.  Precisely, it shows the spectral density  $|V_{s}(\Omega) |^2  = |V_{i}(-\Omega) |^2  $, such that 
\beq 
\left\langle  \hat{A}_s^{\dagger \text{out}}(\Omega)  \hat{A}_s^{\text{out}}(\Omega')  \right \rangle 
= \left\langle  \hat{A}_i^{\dagger \text{out}}(-\Omega)  \hat{A}_i^{\text{out}}(-\Omega')  \right \rangle 
= \delta (\Om- \Om ' ) |V_{s}(\Omega) |^2
\label{Vs}
\eeq
Notice that the singular Dirac $\delta$ in this equation is an artifact coming from assuming a monochromatic pump, of infinite duration, and would  disappear with a proper regularization. 
The plots in Fig.\ref{fig3_spectra} have been calculated  for a periodically poled Lithium Niobate (PPLN)  slab of 1 cm length, pumped at $\lambda_p =771$nm,  with a poling period $\Lambda = 212.7$nm, suitable to phase match the degenerate type 0 interaction at $\lambda_s= \lambda_i = 1542$nm. The wave-numbers were evaluated using the complete Sellmeier relations in \cite{Nikogosi͡an2005}. Actually, these plots as a function of the normalized frequency are approximately valid for any material and tuning condition of the MOPO, because in the linear approximation   \eqref{linear} the spectra are just function of $\Omega/\OmGVS$   As  described in detail in Ref.\cite{Corti2016},  they show  that well below the MOPO threshold parametric generation occurs   in the narrow bandwidth $\OmGVS$, and that on approaching threshold the bandwidth shrinks [fig.\ref{fig3_spectra}a)], while the mean photon number grows[fig.\ref{fig3_spectra}b)], ideally to infinite\cite{Corti2016}.
\par 

On the other side, an analogous  Taylor expansion of the propagation phase  $\beta(\Om) $ in  Eq. (\ref{beta}) gives
\begin{align} 
\beta(\Omega) &= (k'_s-k'_i)\frac{l_c}{2}\Omega  +\frac{l_c}{4}(k_s'' + k_i'')\Omega^2+\cdots \\
&= 
\dtA \Omega  + \tauGVD^2 \Om^2 +\cdots
\label{linear_beta}
\end{align} 
where 
\begin{align}
\dtA &=\frac{1}{2} \left[\frac{l_c}{v_{gs}}-\frac{l_c}{ v_{gi}} \right] \equiv  \OmGVM^{-1}
\label{dtA} \\
\tauGVD &= \sqrt{  \frac{l_c}{4}(k_s'' + k_i'')} \equiv  \OmGVD^{-1}
\label{tauGVD}
\end{align}
are { short} time scales, linked respectively to the group velocity mismatch between the two waves and to the dispersion of the group velocities. Notice that these two time scales are also characteristic of the co-propagating interactions, and are short compared to $\tauGVS$ . 
Taking the example of  the PPLN of Fig. \ref{fig3_spectra} we have: 
 $ \tauGVS= 74.9 \text{ps}$  ($\OmGVS= 1.336 \times 10^{10} \text{rad s}^{-1}$) ;  $\tauGVM=0$; 
$ \tauGVD= 0.016\text{ps}$  ($\OmGVD= 4.45 \times 10^{13} \text{rad s}^{-1}$). 
As a consequence,  the  associated bandwidth are broad, as compared to the MOPO bandwidth $\OmGVS$, and the phase $\beta(\Om)$ has a very slow variation inside the whole bandwidth of emission, a peculiar circumstance of the counterpropagating interaction, that as we shall see has important consequences on the squeezing.

\section{General properties of the output state}
\label{sec:state}
Several properties of  the quantum state of the MOPO below threshold are common 
 to any PDC process, and  in general to all linear  processes where particles are generated in pairs, because they depend  solely on the  form of the Bogoliubov  input -output transformation \eqref{inout}. This section will review these properties,  and  outline their link to the  coefficients  of the transformation \eqref{inout}.  
\par
In order to avoid  formal difficulties coming to the continuum of  modes, in the following of this section we assume some form of discretization of frequencies, e.g by taking a finite quantization time interval $T$, so that  $  \Om \to  \Om_n =\frac{2\pi n }{T}  $  becomes  a discrete set of frequencies . 
\par
In the monochromatic pump approximation,  parametric coupling exists only between  frequency-conjugate spectral components of the signal and idler fields, say 
 $\A_s (\Om)$  and  $\A_i (-\Om) $ \footnote{Notice that for a finite pump bandwidth, coupling will exists in a range of frequencies proportional to the pump bandwidth}, as expressed by the Bogolubov transformation \eqref{inout}
The unitarity of such  transformation  constrains its coefficients $U_j(\Om)$ $V_j(\Om)$ to obey the   conditions \eqref{unitarity}, so that they can be recast   in terms of fewer parameters. In particular, by introducing the {\em squeezing parameter}  $\rrOm$ and the 
{\em squeezing angle} $\theta (\Om)$
\begin {align} 
&|U_s(\Om)| = |U_i(-\Om)|: = \cosh {[r(\Om)]}  \\
&|V_s(\Om)| = |V_i(-\Om)|: = \sinh {[r(\Om)]}  \label{erre}\\
 &\text{arg}\left[  U_s(\Om) V_i(-\Om) \right] =  \text{arg}\left[  U_i(-\Om) V_s(\Om) \right]: = 2\theta (\Om) \, ,
\label{theta}
\end{align}
the input-output transformation can  be rewritten  in the standard form of a {\em two-mode squeeze}  transformation: 
\bsub
\begin{align}
& \hat{A}_s^{\text{out}}(\Om)= \cosh{ [r (\Om)] }  \hat{ B}_s^{\text{in}}(\Omega)
+ e^{2i \theta (\Om) }  \sinh{ [r (\Om)] }\hat{B}_i^{\text{in} \dagger}(-\Omega)\\
&\hat{A}_i^{\text{out}}(-\Om)=   \cosh{ [r (\Om)] }    \hat{B}_i^{\text{in}}(-\Omega)
+e^{2i \theta (\Om) }  \sinh{ [r (\Om)] } \hat{B}_s^{\text{in}\dagger}(\Omega) \, 
\end{align}
\label{inout2}
\esub
where the  new input operators are are just phase rotated versions of the original input operators:   $\hat B_s^{\mathrm in} (\Om) =
e^{i \varphi_{us}  (\Om) }  \hat A_s^{\mathrm in} (\Om)$, $ \hat B_i^{\mathrm in} (-\Om) = e^{i \varphi_{ui} (-\Om)  }  \hat A_i^{\mathrm in} (-\Om)$, with $ \varphi_{uj} (\Om) = \text{arg}[ U_j (\Om) ] $.
 Clearly, such a rotation has no effect on the input vacuum state.
The generator of the transformation \eqref{inout2} is the {\em two-mode squeeze} operator (see e.g Refs. \cite{Knight2005a}, \cite{Gatti2003})
\begin{align}
& \hat R (\xi)=e^ {\sum_{\Om} \left[ \xi (\Om) \A_s^\dagger (\Om ) A_i^\dagger (-\Om) - \xi^* (\Om)  \A_s (\Om ) A_i(-\Om)\right] }\, , &  &
\xi (\Om) = \rr (\Om) e^{2i \theta(\Om) }  \, ,
\end{align}
such that  Eq. \eqref{inout2} can  be recast as: 
\bsub
\begin{align}
& \hat{A}_s^{\text{out}}(\Om)=  \hat R^\dagger (\xi) \hat{ B}_s^{\text{in}}(\Omega) \hat R  (\xi) 
\\
&\hat{A}_i^{\text{out}}(-\Om)=   \hat R^\dagger (\xi)   \hat{B}_i^{\text{in}}(-\Omega)\hat R  (\xi) 
\label{inout3}
\end{align} 
\esub
If the same transformation,  instead of acting on the input  operators,  is applied to the input vacuum state, it generates  at the output  of the crystal the entangled state of twin beams. This  can  be written as the tensor product of states belonging to  subspaces at fixed  $\Om$ : 
\beq 
| \Psi  \rangle^{\mathrm out}  = \prod_{\Om} | \psi \rangle^{\mathrm out} _\Om\, ,
\eeq 
where $| \psi \rangle^{\mathrm out} _\Om $ indicates  the state of  the two coupled signal and idler  modes at  frequencies $\omega_s +\Om$  and   $\omega_i -\Om$.  It can be calculated as 
\begin{align}
| \psi  \rangle^{\mathrm out}_\Om &= \hat R (\xi) |0\rangle_{\Om, s} \, |0\rangle_{-\Om,i} 
= \sum_{N=0}^{+\infty} c_N (\Om) |N\rangle_{\Om, s} \, |N\rangle_{-\Om,i} \, , \\
c_N (\Om) &= \frac {  \tanh [\rrOm]^N  } {\cosh [\rrOm]}   e^{2i N \theta (\Om) }\, , 
\end{align} 
where $ |N\rangle_{\Om, s}$, and $|N\rangle_{-\Om,i} $ denote Fock states,  with N photons  in each of the two modes. This is the well known {\em two-mode squeeze state}, which is an eigenstate with null eigenvalue of the difference of the signal/idler photon numbers at $\pm \Om$ : 
\beq
\left[ \hat A^\dagger_s (\Om) \hat A_s (\Om)- \hat A^\dagger_i (-\Om) \hat A_i (-\Om)\right]  | \psi  \rangle^{\mathrm out}_\Om= 0 \, .
\eeq
This implies the existence of  a perfect correlation between the  photon numbers detected at each pair of  conjugate frequencies of the  twin beams.  
In Sec. \ref{sec:intensity} we shall come back to a more operative definition of the photon number correlation in the spectro-temporal  continuum. 
\par
Besides the photon-number correlation, this state is also well known to 
to display a noteworthly EPR-type of  correlation between two non-commuting  quadrature operators of the signal and idler field \cite{Reid1989, Ou1992}. 
Precisely,  if one  focuses on a pair of frequency-conjugate spectral components  $\omega_s +\Om$, $\omega_i -\Om$ , and introduces their   sum and difference
\begin{align}
\hat c_{\pm} (\Om) &= \frac{\hat A_s^{\mathrm out} (\Om) \pm  A_i^{\mathrm out} (-\Om)}   {\sqrt 2}, 
\label{sumdiff}
\end{align} 
then  
the transformation \eqref{inout} decouples into two independent {\em squeeze} transformations 
\bsub
\begin{align}
& \hat c_+ (\Om)= \cosh{ [r (\Om)] }  \hat{ B}_+^{\text{in}}(\Omega)
+ e^{2i \theta (\Om) }  \sinh{ [r (\Om)] }\hat{B}_+^{\text{in} \dagger}(\Omega)\\
&\hat c_- (\Om)=   \cosh{ [r (\Om)] }    \hat{B}_-^{\text{in}}(\Omega)
- e^{2i \theta (\Om) }  \sinh{ [r (\Om)] } \hat{B}_-^{\text{in}\dagger}(\Omega).
\end{align}
\label{inout4}
\esub
where $ \hat{ B}_{\pm}^{\text{in}}$ are independent modes in the vacuum state, defined in obvious way as the sum and difference of the input signal-idler modes. 
Thus the  $\pm$ modes, which combine frequency conjugate signal and idler spectral components (notice that  this slightly differs to what is in practice done in a measurement, see next section)  are independent and individually squeezed.   The two parameters $\theta$ and $\rr$, defined by Eq.\eqref{erre},  determine the phase angle at which noise reduction occurs, and the maximum level of squeezing achievable, respectively. Namely, for the difference  mode $\hat c_- (\Om) $,  best  squeezing  occurs for the field quadrature at angle $\theta (\Om) $, while for the sum mode   $\hat c_+ (\Om) $ it occurs in  the orthogonal quadrature at angle $\theta (\Om)+ \pi/2 $.   
At these angles,  quantum noise is reduced below the shot noise value "1" by an amount 
\begin{align} 
e^{-2\rrOm  }  &= \left| U_s(\Om) \right|  - \left| V_i(-\Om)  \right|^2
= \frac{1}{  \left| U_s(\Om)  \right| + \left| V_i(-\Om)  \right|^2   }
\label{noisereduction} 
\end{align} 

 As well know, this means that the signal and idler quadratures at angle $\theta (\Om) $ are correlated, while at the same time  the orthogonal  quadratures at angle $\theta (\Om) + \pi/2 $ are anticorrelated. The degree of simultaneous correlation/anticorrelation in the orthogonal quadratures can be large enough to provide a realization of the original EPR paradox \cite{Reid1989, Ou1992}.

\section{Quadrature correlation in the MOPO below threshold}
\label{sec:quad}
In order to characterize the squeezing and EPR correlation generated in specific case of the MOPO, let us come back to the continuum of frequencies and introduce  
 a definition of the phase sensitive noise suitable for measurements. To this end, we consider the  signal ($j=s$ ) and idler ($j=i$)quadrature operators in the time domain: 
\bsub
\begin{align}
\hat X_j (t) &= \hat A_j^ {\mathrm out}  (t)e^{-i \phi_j } + \hat A_j^ {\mathrm out \, \dagger }  (t)e^{i \phi_j} 
\label {Xtime} ,  \\
\hat Y_j (t) &= -i \left[ \hat A_j^ {\mathrm out} (t)  e^{-i \phi_j} - \hat A_j^ {\mathrm out \, \dagger }  (t)e^{i \phi_j}\right]
\label{Ytime}
\end{align}
\label{Qtime}
\esub
which, by varying the phase-angles $\phi_j$, span all the classical phase-space of the harmonic oscillator, remaining orthogonal $ \hat X \to \hat Y$ for $\phi_j \to \phi_j + \frac{\pi}{2 }$. 
Next, we  introduce proper combinations of these signal and idler operators: 
\begin{align}
&\hat X_{-} (t) = \frac{ \hat X_s(t) -\hat X_i (t ) }{\sqrt 2} & , 
\hat Y_+ (t) = \frac{ \hat Y_s(t) + \hat Y_i (t ) }{\sqrt 2} ,
\label{Xpm_time}
\end{align} 
which basically represent the quadrature operators of the modes $\hat c_{-} $, $\hat c_{+} $  in Eq. \eqref{sumdiff}. 
The 	quadrature noise in the sum and difference modes is then characterized by the {\em spectra of squeezing} 
\bsub
\begin{align}
\Sigma_{-} (\Omega) &= \int_{-\infty} ^{+ \infty}  d\tau \, e^{i \Om \tau}  \left\langle \delta \hat X_-(t) \hat \delta X_-(t + \tau) \right\rangle  \\
\Sigma_{+} (\Omega) &= \int_{-\infty} ^{+ \infty}  d\tau \, e^{i \Om \tau}  \left\langle \delta \hat Y_+(t) \hat \delta Y_+(t + \tau) \right\rangle
\end{align}
\label{Sigmapm}
\esub
These quantities describe the degree of  correlation ("-" sign) or anticorrelation ("+" sign) existing between the field quadrature operators of the twin beams at the two crystal output faces.  With our definitions, the value "1" represents the shot noise level, which corresponds to two uncorrelated coherent light beams. In the case of  signal and idler fields with the same central frequency, one may also think of physically 
recombining  the two counterpropagating light beams,  in order to produce two independently squeezed  beams. \\
After some calculations, based on the input-output relations \eqref{inout}, we obtain
\begin{align}
\Sigma_{\pm} (\Om )  &= \frac{1}{2}  \left[ {\cal F} (\Om) + {\cal F} (-\Om)  \right] \nn \\
{\cal F} (\Om) &=  \left|  U_s(\Om)  - V_i^* (-\Om)  e^{i (\phi_s + \phi_i)} 
\right|^2  
\label{Sigmapm2}
\end{align} 
The two symmetric spectral components ${\cal F} (\Om)$, ${\cal F} (-\Om)$ 
 can be shown to represent  the noise  in  the two sidebands modes $ \hat c_{\pm}  ( \Om)$, and $ \hat c_{\pm}  ( -\Om)$ \cite{Gatti2017} .
As described in \cite{Gatti2017}, the best squeezing is achieved by choosing the phase-angles as 
\begin{align} 
\phi_s + \phi_i    = 2\theta (\pm \Om) &= k_s l_c + \phi_p   +\text{arg}\left[  \sinc \gamma (\pm \Om) \right]  + \beta (\pm \Om)
\label{phiopt1} \\
&\simeq   k_s l_c + \phi_p   +\text{arg}\left[  \sinc \gamma (\Om) \right]  
\label{phiopt}
\end{align}  
where  the last line uses the fact that the phase $\beta $ is almost constant over the entire MOPO spectrum, and the  linear approximations \eqref{linear}, which implies $\gamma (\Om)= \sqrt{g^2 + \frac{\Om^2}{\Omega^2_{gvs}}}
= \gamma (-\Om)$.  \footnote{Alternatively, as explained in Ref.\cite{Gatti2017} in the presence of GVM, a further optimization can be done by introducing   a delay  $\Delta  t = \dtA$  between the detection of the twin beams, which  compensate the  offset between their exit times.}
\begin{figure}[ht]
     \includegraphics[width=0.9\textwidth]{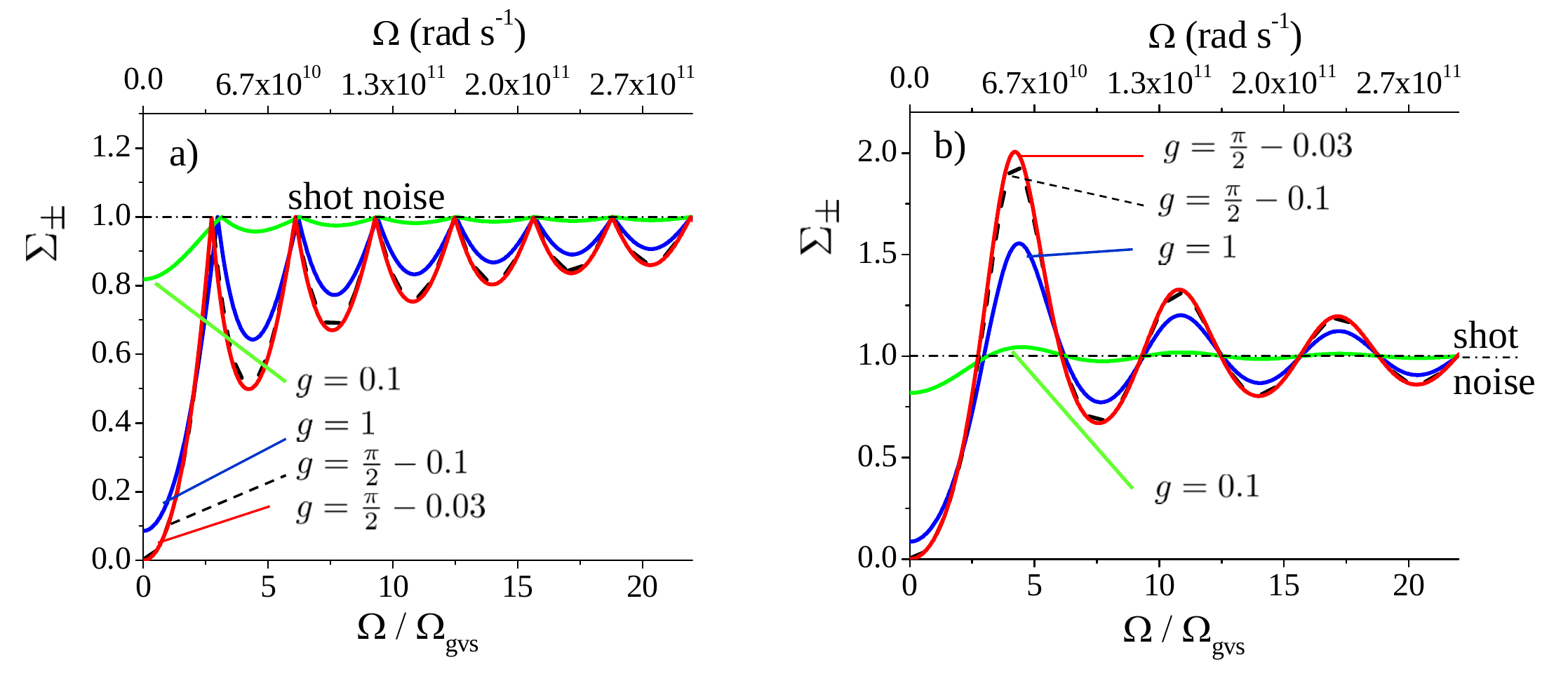}
           \caption{(Color online) Squeezing spectra $\Sigma_{\pm} (\Omega) $  \eqref{Sigmapm} in the sum or difference modes, and degree of EPR correlation between field quadratures of the MOPO twin beams,plotted for different values of the gain $g$,  as a function of  the normalized frequency $\Omega/\OmGVS$ (bottom axis) and of the frequency (upper axis) in a PPLN counterpropagating slab.   In  a) the quadrature angles  are optimized for best squeezing  [Eq.\eqref{phiopt}]. In b)  they are fixed as $\phi_s + \phi_i = k_sl_c + \phi_p $ . Parameters as in Fig.\ref{fig3_spectra}  }
\label{fig_squeezing}
\end{figure}
With this choice
\beq
\Sigma_{\pm} (\Om ) \to  \left[  \left| U_s(\Om) \right| -\left| V_i (-\Om)\right| \right]^2=e^{-2r(\Om)}
\label{Sigmaopt}
\eeq
reaches its minimum value at any frequency, and the noise never goes above the shot noise level "1", as shown by 
Fig.\ref{fig_squeezing}a. 
Fig.\ref{fig_squeezing}b shows instead the degree of squeezing/EPR correlation when the phase-angles  are fixed as 
\beq
\phi_s + \phi_i  =  2\theta (0) =  k_s l_c + \phi_p  
\label{phifix}
\eeq
As described in detail Ref. \cite{Gatti2017},  we see that when the MOPO threshold is approached the level of squeezing and EPR correlation become asymptotically perfect, showing a behaviour  completely analogous to that of a cavity OPO below threshold, which is the standard source of squeezed light. Good  levels  of EPR correlation are present  even quite far from threshold, having for example a $90\%$ squeezing for $g=1$, which is 36$\%$ below the threshold. 
\par
Most important,  as evident from a comparison between  Fig.\ref{fig_squeezing}b and the MOPO spectra in Fig. \ref{fig3_spectra}, excellent levels of squeezing and EPR correlation can  be obtained inside the entire emission bandwidth of the MOPO even when the phase angles are fixed as in  Eq. \eqref{phifix} $\phi_s + \phi_i = 2\theta(0)$.   This is important because in practice the detection will be performed mostly at  fixed phase angles. 
In this case,  the noise passes from below to above the shot noise at  $|\Omega| = \OmGVS \sqrt{\pi^2 -g^2} \,  $, i.e. at the point where the $\sinc {\gamma(\Om)}$ changes sign. Thus   the bandwidth of squeezing $\Delta \Omega_{\mathrm squeeze} = 
\OmGVS\sqrt{\pi^2 -g^2} $ remains  approximately constant in the neighborhood of the threshold  
$\Delta \Omega_{\mathrm squeeze} \approx 2.7 \OmGVS$.
\par
\begin{figure}[ht]
     \includegraphics[width=0.9\textwidth]{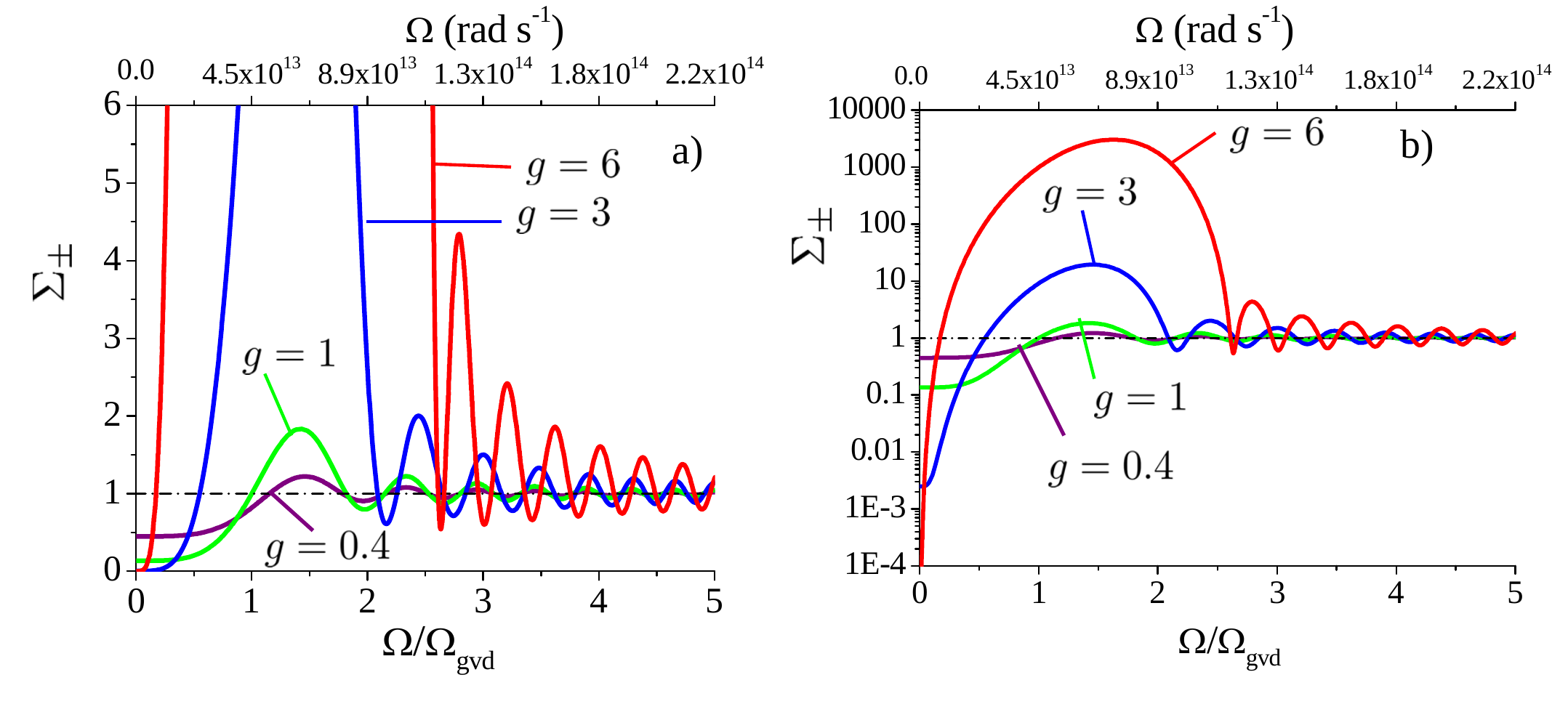}
           \caption{(Color online) Co-propagating configuration: a) Squeezing spectra $\Sigma_{\pm} (\Omega) $  \eqref{Sigmapm}, plotted for different values of the gain $g$,  as a function of  the normalized frequency $\Omega/\OmGVD$ (bottom axis) and of the frequency (upper axis) in a PPLN co-propagating slab.   The quadrature angles are fixed as $\phi_s + \phi_i = \theta(0) = ( k_p -k_G) l_c + \phi_p $ .    b) Same plot in logarithmic scale.  $\lambda_p =771$nm,   $\lambda_s= \lambda_i = 1542$nm, $\Lambda = 18.8 \mu$m, $l_c=1$cm.}
\label{squeezing_coprop}
\end{figure}

\begin{figure}[ht]
     \includegraphics[width=0.9\textwidth]{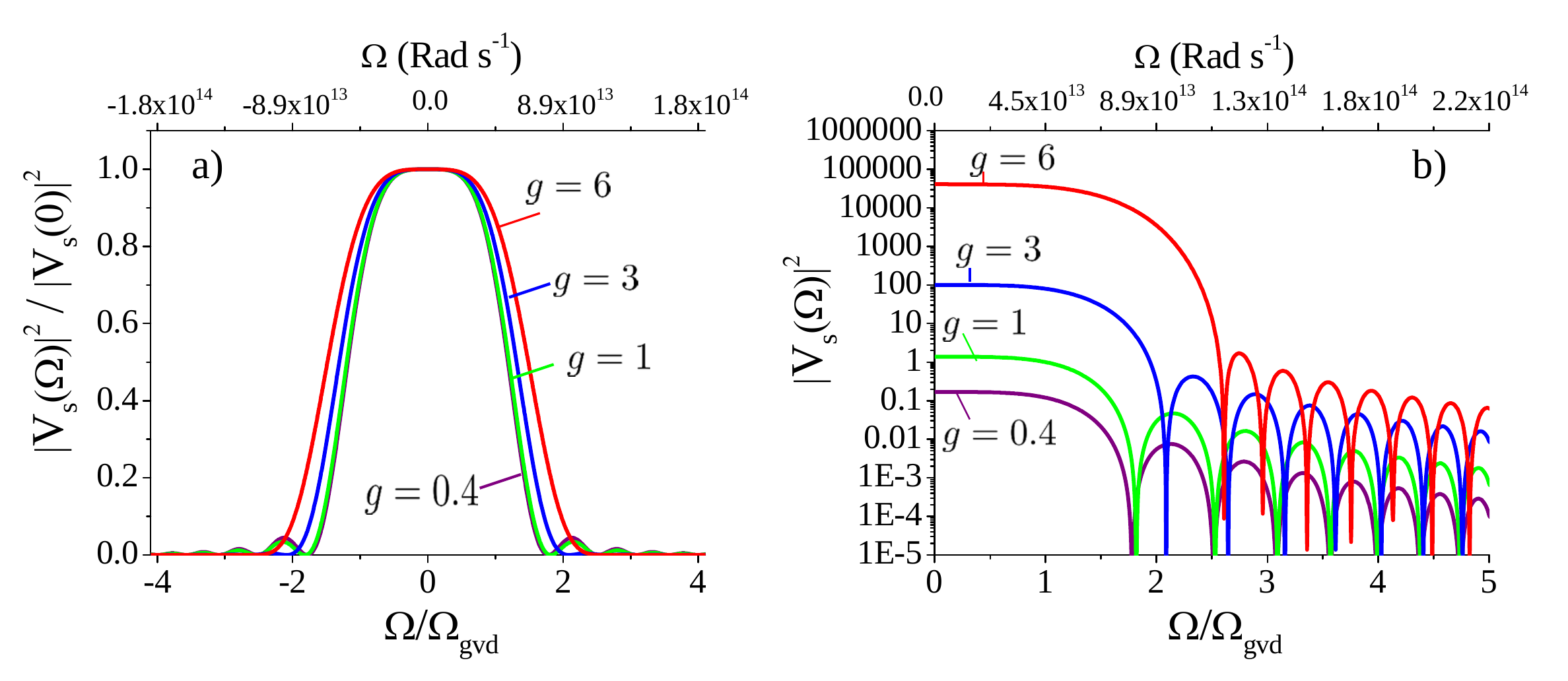}
           \caption{(Color online)Co-propagating configuration:  Intensity spectra $|V_{s}(\Omega) |^2    $  \eqref{Vs}, plotted for different values of the gain $g$, as a function of  $\Omega/\OmGVD$ (bottom axis) and of $\Om$ (upper axis) in a PPLN slab  poled for the co-propagating type 0 interaction. Parameters as in figure \ref{squeezing_coprop} }
\label{spectra_coprop}
\end{figure}

 
It is interesting to compare this behavior with what can be obtained in the standard single-pass co-propagating configuration.  Within the monochromatic and undepleted pump approximations this setup can be  modeled by the same  input-output Bogoliubov tranformation \eqref{inout}, with the obvious difference that   the output field operators appear in this case on the same face of the slab. The  coefficients $U_j$ and $V_j$ have to be substituted by those calculated for the co-propagating parametric equations, and can be for example found in Refs. \cite{Gatti2003,Brambilla2004} (the transverse wave-vector  appearing there has to be set to $\vec q= 0$ in order to describe collinear propagation). 
The squeezing spectra and the intensity spectra are then calculated from the same Eqs. \eqref{Sigmapm} and \eqref{Vs}, with the proper coefficients inserted. We can take the example of a 1 cm  PPLNslab, pumped at 771 nm, similarly to the MOPO case, but with  a long poling   poling period  $\Lambda= 18.8 \mu$m, chosen  to phase match the co-propagating degenerate type 0 interaction.  Results for the squeezing spectra and for the intensity spectra are shown in figures \ref{squeezing_coprop} and \ref{spectra_coprop}, respectively.  
The relevant spectral scale is in this case the ultrabroad dispersion bandwidth $\OmGVD$, because the phase mismatch for the co-propagating configuration is given by 
\begin{align}
\Delta (\Om) \frac{l_c}{2} &= \left[ k_s(\Om) + k_i (-\Om) - k_p +k_G \right]\frac{l_c}{2}
= \left( k_s'  - k_i ' \right)   \frac{l_c}{2} \Om +  \frac{k''_s + k_i"}{2} \frac{l_c}{2} \Om^2 + ... \nn \\
&= \tauGVM \Om + \tauGVD^2 \Om^2 \to \left (\frac{\Om}{\OmGVD}\right)^2
\end{align}
where the last result holds for the  type 0 or type I interactions at degeneracy, where $\tauGVM=0$. The parameters $\tauGVM, \tauGVD$ and $\OmGVD$ are the same as defined by Eqs. \eqref{dtA}  and \eqref{tauGVD}. 
\\
As  well known,  the fluorescence spectra are in this case ultrabroadband [see Fig. \ref{spectra_coprop}],  because ruled by the dispersion bandwidth 
$\OmGVD$, and tend to become slightly broader when the parametric gain increases (notice that $g$ in these figure is defined by  Eq. \eqref{gain}, exactly in the same way as for the MOPO, but obviously there is no threshold value).  In contrast, the squeezing spectra in Fig.\ref{squeezing_coprop}  exhibit noise reduction below the shot noise in a smaller bandwidth, which in particular  shrinks dramatically as soon as the parametric gain grows above the value $g \simeq 1$ at which  stimulated emission starts to be important. 
Then, at high gain,   it would be difficult to observe squeezing in the whole fluorescence  bandwidth. This happens because  the orientation  of the squeezing ellipses  varies rapidly with the frequency inside the broad PDC bandwidth. Calculations based on the explicit solution of the co-propagating parametric equations in \cite{Gatti2003,Brambilla2004} show that in this case the angle at which best squeezing occurs is 
$\theta (\Om) -\theta(0) \simeq \left (\frac{\Om}{\OmGVD}\right)^2 \frac{\tanh g}{g} $, thus varying   on a similar scale $ \simeq \OmGVD$ as the spectrum.  Notice that, when the phase-angles are fixed as $\phi_s + \phi_i = 2 \theta (0)$,  the detected squeezing spectra can be written as: 
\begin{align}
\Sigma_\pm (\Om) &= \left| U_s(\Om) - V_i^* (-\Om) e^{2i  \theta(0)}\right|^2  
= 
e^{-2 \rr (\Om)}  + 2 \sinh{[2 \rr (\Om) ]} \sin^2{ [\theta(\Om)-\theta(0)] } \nn  \\
&\approx 
e^{-2 \rr (\Om)}  +   e^{2 \rr (\Om)}\sin^2{ [\theta(\Om)-\theta(0)] } \qquad \text{for } \rr (\Om) \gg 1 
\end{align} 
For significant squeezing $\rr \ge 1$,   as soon as the squeezing phase has a small change with respect to $\theta(0)$,  a large excess noise (the second term at r.h.s) appears and degrades the squeezing. 
\\
This behaviour  is
 in sharp contrast with 
 the MOPO,  where the  orientation of the squeezing ellipses, defined by Eqs.(\ref{phiopt1}-\ref{phiopt})  varies slowly  with  the frequency, 
and  remains constant  inside the narrow MOPO bandwidth  $ \simeq \OmGVS$. This can be seen  as a consequence of the  peculiar time scale $\tauGVS$ involved  in the counterpropagating interaction, which is much longer than the other time scales $\tauGVM, \tauGVD$ that  characterize the co-propagation of light waves. 
\par
Notice that  in the co-propagating case the bandwidth of   squeezing,   despite it shrinks with increasing gain, remains  in any case ultrabroad, in the Thz region (see the upper scales in Figs. 4 and 5), and probably  could  be  further enlarged by properly shaping the phase of the local oscillator. 
\section{Intensity correlation}
\label{sec:intensity}
We now turn our attention to the  quantum correlation between the intensities of the MOPO twin beams.  
\par 
As a trivial consequence  of the pairwise generation of photons, when the MOPO intensities   are collected   for a long enough time, the detected photon numbers will be identical not only in the mean values, but also in their quantum fluctuations.  
A good question is then how long one has to detect before the photon-number difference shows fluctuations below the shot-noise that  represents the classical limit? Our curiosity in this sense is driven by the findings of Ref. \cite{Corti2016}. Here it was shown that in the spontaneous regime the MOPO twin photons are correlated over the time $ \tau_{corr}= \tauGVS$, which simply reflects the fact 
that  twin photons originating from the same pump photon exit the crystal  at most delayed by their transit time across the slab. 
In the stimulated regime, however,  the correlation becomes long ranged, and ideally $\tau_{corr} \to \infty$ as threshold is approached, as a consequence of a combination of  stimulated emission and back-propagation.  Then,  we ask ourselves whether close to threshold an infinite detection time would be needed to observe sub-shot noise intensity  fluctuations. 
\par 
To this end, we consider the instantaneous intensity operators  at the  crystal output faces 
$
\hat  I_j(t)=\hat A_j^{\dagger\mathrm{out} }(t)\hat A^{\mathrm{out}}_j(t)$ $(j=s,i )$.
In order to describe their level of  correlation, on the one side we introduce  the intensity  difference 
\beq
\hat  I_-(t)=\hat  I_s(t)-\hat  I_i(t) \, 
\label{Imeno}
\eeq
and calculate  its   spectrum of  fluctuations: 
\beq
{\cal V}_-(\Omega)=\int d\tau e^{i\Omega \tau} \braket{\delta I_-(t)\delta I_-(t+\tau)} .
\label{Vmeno}
\eeq
On the other side, we also consider 
 the photon-number operators  that result from integrating the intensities over a finite detection time 
${T_d}$: 
\beqa
\hat N_j &=&\int_{-\frac{T_d}{2}}^{\frac{T_d}{2}} dt \hat I_j (t)=\int_{-\frac{T_d}{2}}^{\frac{T_d}{2}} dt  
\hat A_j^{\dagger\mathrm{out} }(t)\hat A^{\mathrm{out}}_j(t) \qquad j=s,i  
\label{Enne}
\eeqa
and evaluate the noise in their difference   $ \hat N_-=\hat N_s-\hat N_i$. 
\beq 
\braket{(\delta \hat N_-)^2} = \braket{  (\delta \hat N_s -\delta \hat N_i )^2      }
\label{dNmeno}
\eeq
When these two quantities go below their shot-noise level, characterizing coherent light or classically correlated beams (e.g.  generated by splitting thermal light on a beam splitter\cite{Gatti2004a}), we can then talk of a quantum correlation of microscopic nature. \\
The two approaches are linked, because the variance \eqref{dNmeno} 
can  be written in terms of the 
spectrum (\ref{Vmeno}) as 
\beqa
\braket{(\delta \hat N_-)^2}
&=&
\int_{\frac{-T_d}{2}}^{\frac{T_d}{2}}dt  
\int_{\frac{-T_d}{2}}^{\frac{T_d}{2}}dt' \braket{\delta \hat I_-(t)\delta \hat  I_-(t') } 
\label{dNmeno1} \\
&=& T_d^2\int \frac{d\Omega}{2\pi}\sinc^2\left(\frac{T_d\Omega}{2}\right){\cal V}_-(\Omega) 
\label{dNmeno2}
\eeqa
For an infinite  detection time $T_d \to \infty$, $\braket{(\delta N_-)^2} \to  T_d {\cal V}_-(\Omega=0)=0$. The same  holds true when the detection time largely exceeds the inverse bandwidth of ${\cal V}_-(\Omega) $, because in that case  the  $\sinc$ function behaves as a $\delta$-function under the integral. Thus we expect that the variance \eqref{dNmeno} approaches zero for  detection time $T_d$ longer than the inverse of the bandwidth of the spectrum \eqref{Vmeno}. 
\par
We start by calculating the spectrum of $\hat I_ -$.  To this end, we express  the temporal correlation function of $I_-(t)$  in terms
of the self- and cross-correlation of the intensities of the two fields
\beqa
\braket{\delta \hat I_-(t)\delta \hat  I_-(t+\tau)} 
&= G^{(2)}_{ss}(t,t+\tau)+G^{(2)}_{ii}(t,t+\tau)  
-G^{(2)}_{si}(t,t+\tau)-G^{(2)}_{si}(t,t-\tau) \, ,
\label{Gmeno}
\eeqa
where
\beq
G^{(2)}_{jl}(t,t'):=\braket{\delta\hat I_j(t)\delta\hat  I_l(t')} \qquad j,l=i,s \, .
\eeq
In  Eq.(\ref{Gmeno}) we used the indentity $G^{(2)}_{is}(t,t')=G^{(2)}_{si}(t',t)$,  and 
the fact that $G^{(2)}_{jl}(t,t')$ depends only on the time difference $t-t'$
under stationary conditions. Since the model is linear, the fourth order field moments contained in the   correlation functions $G^{(2)}_{jl}$ can be factorized into second-order field moments 
according to  (see e.g. \cite{Gardiner1991})
\beqa
G^{(2)}_{jj}(t,t') &=& \braket{\delta \hat  I_j(t)\delta \hat I_j(t')}
            =\delta(t-t')\braket{I_j}+\left|   \braket{\hat A_j^{\dagger \text{ out}}(t)\hat A^{\text{out}}_j(t')}  \right|^2,
\label{Gjj}\\
G^{(2)}_{si}(t,t')&= &\braket{\delta \hat  I_s(t)\delta \hat I_i(t') } 
     =\left| \braket{ \hat A_s^{\text{out}}(t) \hat A_i^{\text{out}}(t')} \right|^2\label{Gsi}
\eeqa
The second order field moments can be easily calculated from the input-output relations \eqref{inout}: 
\beqa
\braket{\hat A_s^{\dagger \text{ out}}(t)\hat A_s^{\text{out}}(t')} 
               &=&\int \frac{d\Omega}{2\pi}e^{i\Omega(t-t')}|V_s(\Omega)|^2		
=\braket{\hat A_i^{\dagger \text{ out}}(t')\hat A^{\text{out}}_i(t)} \\
 \braket{ \hat A_s^{\text{out}}(t) \hat A_i^{\text{out}}(t')} 
          &=&\int \frac{d\Omega}{2\pi} e^{-i\Omega(t-t')}U_s(\Omega)V_i(-\Omega)\label{psi}
\eeqa
The first term  at r.h.s. of  Eq.(\ref{Gjj}) represents  the shot-noise contribution, where the signal and idler mean intensities are  
\beqa
\braket{\hat I_s}&=&\int\frac{d\Omega}{2\pi}|V_s(\Omega)|^2=\int\frac{d\Omega}{2\pi}|V_i(\Omega)|^2=\braket{\hat I_i}
\label{Is}
\eeqa
After some manipulations the noise spectrum can then be written as 
\begin{align}
{\cal V}_-(\Omega) = \int  \frac{ d\Om'}{2 \pi} \left| U_s (\Om') V_i^* (-\Om -\Om') 
-U_s (\Om + \Om') V_i^* (-\Om') \right|^2 
\label{Vmeno2}
\end{align} 
This result is general to all processes of photon-pair generation, because it uses only the Bogoliubov transformation \eqref{inout}, 
and shows that ${\cal V_-}$  vanishes identically at zero frequency $\Om=0$. Following the discussion after Eq.\eqref{dNmeno2}, as anticipated,  this implies that  the noise in the photon number difference \eqref{dNmeno} also vanishes for an infinite detection time. 

\begin{figure}[h]
\centering
\includegraphics[scale=.55]{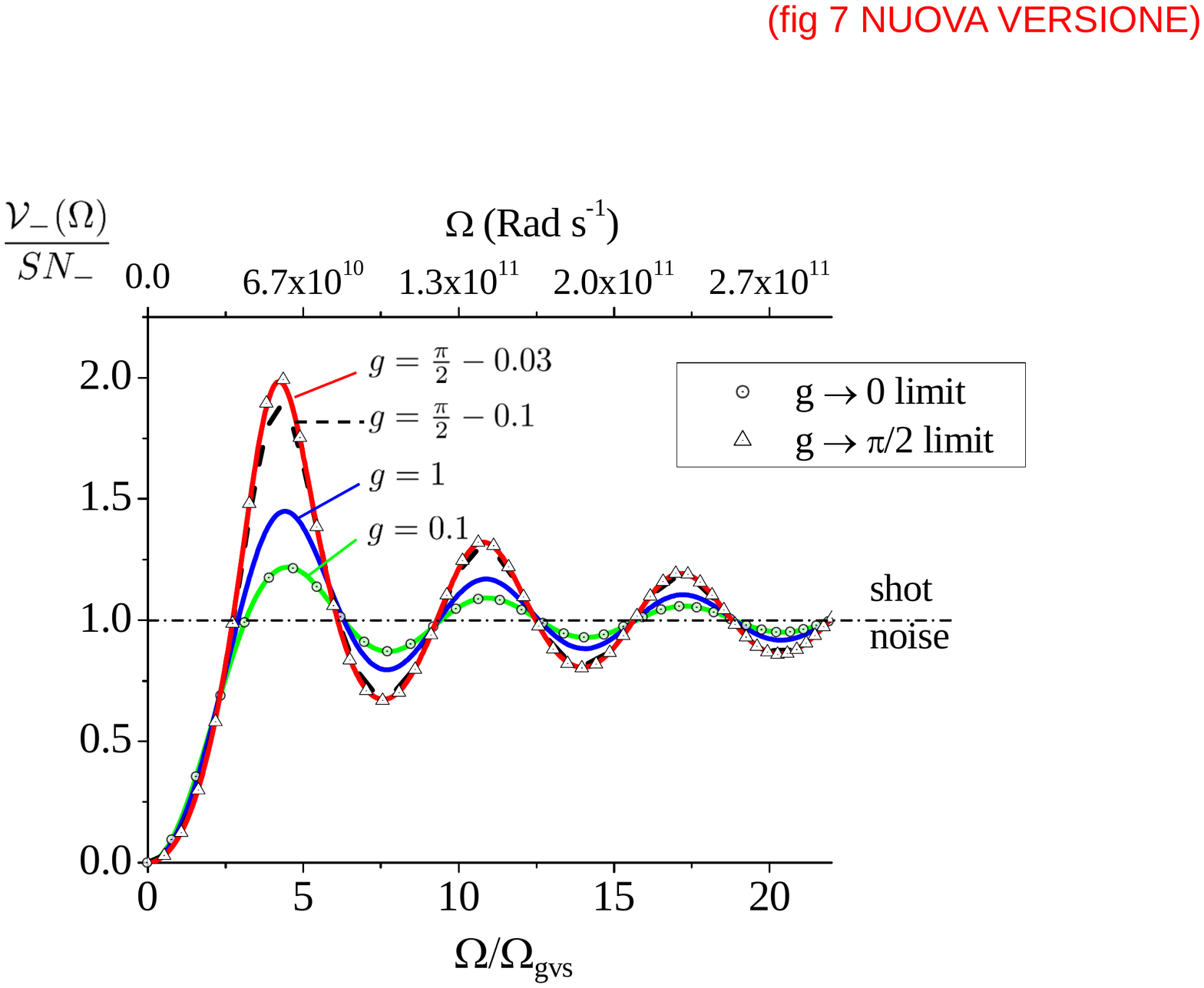}
\caption{Spectrum of $I_-$
normalized to the shot-noise,   in a PPLN counterpropagating slab, as evaluated from numerical integration of
Eq.(\ref{Vmeno3}). The disc and the triangle symbols plot the approximated expressions
 in the spontaneous regime (\ref{Vspont})   and close to threshold  (\ref{Vstim}), respectively. Parameters as in Fig.\ref{fig3_spectra} }
\label{fig_Vnum}
\end{figure}

More insight into the MOPO case can be gained by inserting in Eq. \eqref{Vmeno2} the explicit expression 
of the $U_j,V_j$  [Eq.\eqref{UV}], with which 
the spectrum of $I_-$ becomes 
\begin{align}
{\cal V}_-(\Omega)
=
g^2\int\frac{d\Omega'}{2\pi}   \Big\{   | \phi(\Omega')|^2
|\phi(\Omega+\Omega')|^2  \Big| & \, \sinc[\gamma(\Omega+\Omega')]  +  \phantom{XXXXXXX} \nn \\
&   -  \sinc[\gamma(\Omega')]  
e^{i \beta (\Om+ \Om')} e^{ -i\beta(\Om') }  \Big|^2    \Big\}
\label{Vmeno3}
\end{align}
Figure \ref{fig_Vnum} plots the spectrum ${\cal V}_-(\Omega)$ of the intensity difference, normalized to the shot noise level $SN_-\equiv\braket{I_s}+\braket{I_i}$, 
evaluated through the numerical integration of the exact relation (\ref{Vmeno3}) for different values of the parametric gain $g$. We see that the signal and idler intensities display a sub-shot noise correlation within a bandwidth $\sim 2.5 \,\Omega_{\rm gvs}$
both in the spontaneous and in the stimulated regime.
\par 
Eq.
 (\ref{Vmeno3}) can be used to obtain analytical estimations of the spectrum in the two limiting cases $g\to \frac{\pi}{2}$ (close to threshold) and $g \to 0$ (spontaneous regime) . \\
Close to threshold, for  $g \to \frac{\pi}{2} $ and for values of  $\Omega$ not too far from $\Omega=0$, the function $ |\phi |^2 $ appearing under the integral in Eq.\eqref{Vmeno3} can be approximated as 
\beq
| \phi (\Omega)|^2 \simeq 
\frac{g^2 + \tilde{\Om}^2 }{g^2 \sin^2{\epsilon}  + \tilde \Om^2} \qquad \text{ for }\;\epsilon:=\frac{\pi}{2}-g\to 0
\label{lor}
\eeq
where $\tilde \Om= \frac{\Om}{\OmGVS}$. 
This is a rational function with two simple complex poles  at $\tilde \Om= \pm i g \sin{\epsilon} $, a result that can be used to calculate the integral in Eq-\eqref{Vmeno2} with a contour integration in  the complex plane.  As a result 
\beq
{\cal V}_-  (\Om) \simeq \frac{g\OmGVS }{2\sin{\epsilon} }
\left[ |\phi(\Omega)|^2 \left| e^{i\beta(\Om)} 
-g \sinc [\gamma(\Om)] \right|^2 + \big( \Om \to -\Om\big)  
\right] \qquad \text{for } \epsilon \to 0 
\eeq
An analogous calculation for the shot-noise provides 
\beq 
SN_- = \braket{I_s} + \braket{I_i}  \simeq \frac{g\OmGVS }{\sin{\epsilon} }  \qquad \text{for } \epsilon \to 0 
\eeq 
Both quantities in principle diverge on approaching threshold, but at a finite distance from threshold their ratio gives 
\beqa
\frac{{\cal V}_-(\Omega)}{SN_-}
&\simeq &
\frac{1}{2}\left[ |\phi(\Omega)|^2 \left| e^{i\beta(\Om)} 
-g \sinc [\gamma(\Om)] \right|^2 + \big( \Om \to -\Om\big)  
\right] \qquad \text{for } \epsilon \to 0 
\label{Vstim2}
\\
&= & 
\frac{1}{2}\left[  \left| U_s(\Om) - V_i^*(-\Om) e^{i k_s l_c + i \phi_p} \right|^2 + \big( \Om \to -\Om\big)  
\right]   
\label{Vstim}
\eeqa
Remarkably, the last expression coincides
with the quadrature squeezing spectra $\Sigma_\pm(\Omega)$  in Eq.\eqref{Sigmapm},  evaluated 
at the fixed phase angles $\phi_s + \phi_i = k_s l_c + \phi_p =2 \theta(0)$, as confirmed by a comparison between the spectra in figures \ref{fig_squeezing}b  and \ref{fig_Vnum}. Thus 
\beq
\frac{{\cal V}_-(\Omega)}{SN_-} = \Sigma_\pm(\Omega) \approx
\frac{  \sqrt{g^2 + \tilde{\Om}^2}  - g \sin{ \sqrt{g^2 + \tilde{\Om}^2}}   }
{\sqrt{g^2 + \tilde{\Om}^2}  + g \sin{ \sqrt{g^2 + \tilde{\Om}^2}} }
\qquad \text{for } g\to \frac{\pi}{2}
\label{analytic2}
\eeq
where $\tilde \Om = \Omega/\OmGVS$, and the last formula at r.h.s is obtained by using the linear approximation \eqref{linear} and  setting $\beta (\Om) \simeq 0$\cite{Gatti2017}.
Notice however that, at difference with the quadrature spectra, 
in the region where squeezing is present (say $\Omega \le 2.7 \OmGVS) $   the curve ${\cal V}_- / SN_- $  remains practically identical passing from the highly stimulated regime $g \simeq \pi/2$ to the spontaneous regime $g \ll 1$. 
In the latter case,  for  $g \to 0$,  the function 
$|\phi(\Omega)|^2=1+|V_s(\Omega)|^2 \to 1  $ , and  can be substituted with unity  in Eqs.(\ref{Is}) 
and (\ref{Vmeno2}) (this amounts  to keeping only the leading order
terms in $g^2\ll 1$). Then 
\beqa
\frac{{\cal V}_-(\Omega)}{SN_-} 
&\simeq&
1-\sinc\left(\frac{\Omega}{\Omega_{\rm gvs}}\right)\quad \quad \text{for}\;g\rightarrow 0 
\label{Vspont}
\eeqa

\begin{figure}[h]
\centering
\includegraphics[scale=.8]{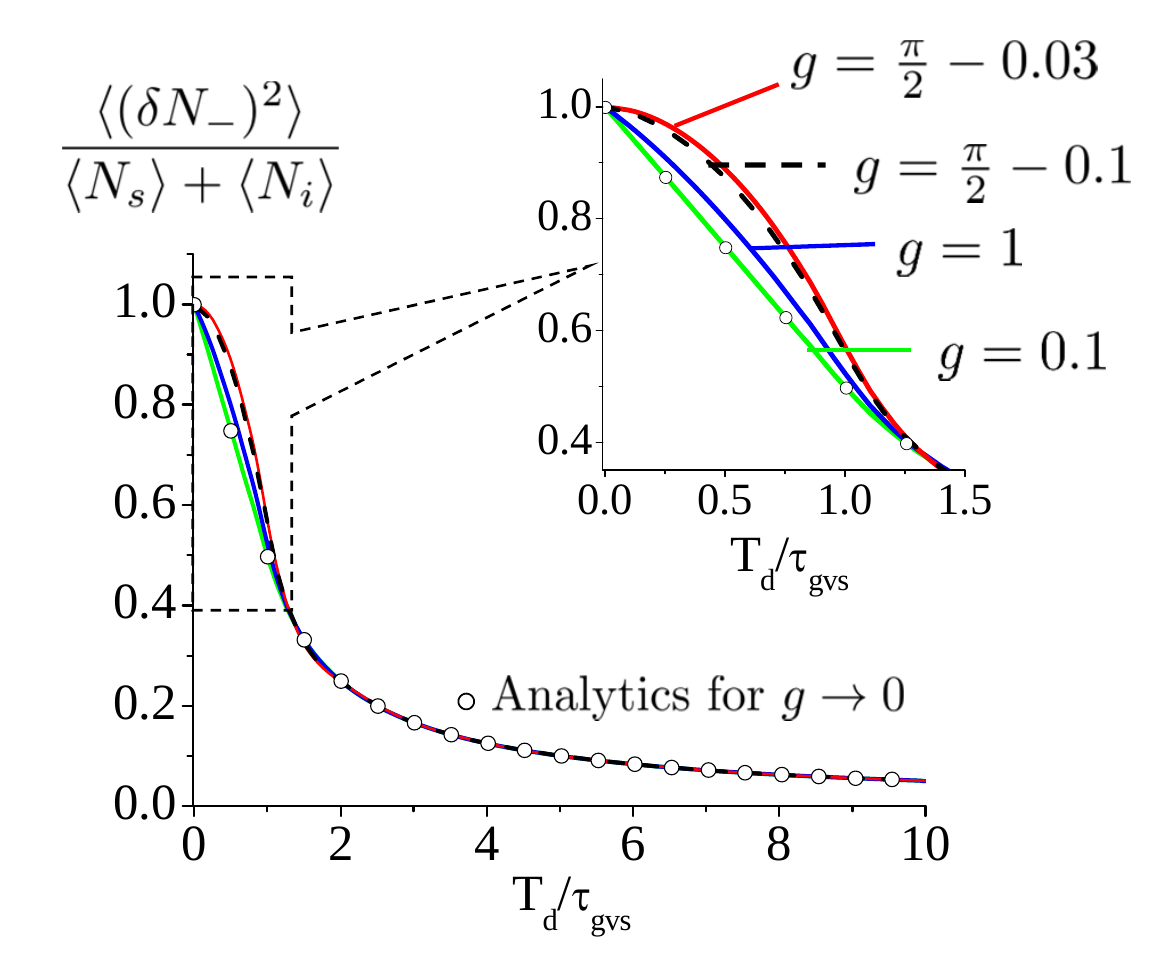}
\caption{Noise in $\hat N_-$ normalized to the shot noise level $\braket{N_s}+\braket{N_i}$ as a function of the detection time $T_d$.
The  formula (\ref{dNspont}) derived in the limit $g \to 0$ (open circles) perfectly fits  the $g=0.1$ result (green line), and   reproduces rather well  the curves in all regimes. Parameters as in Fig.\ref{fig3_spectra} }
\label{fig_Nmeno}
\end{figure}

Finally,  we investigate the noise in the difference of photon numbers collected over a finite time window $T_d$. 
Results are reported in Fig.\ref{fig_Nmeno},
 which  plots $\braket{(\delta N_-)^2}$ divided by the shot noise  $\braket{N_s}+\braket{N_i}$ as a function of $T_d/\tau_{\rm gvs}$ for different values of the parametric gain. \\
In the spontaneous regime (limit $g\rightarrow 0$) it is possible to derive an approximated expression for this noise variance (\ref{Appendix}): 
\beqa
\frac{\braket{(\delta \hat N_-)^2}}{\braket{\hat N_s}+\braket{\hat N_i}}
\simeq
\begin{cases}
1-\frac{T_d}{2\tau_{\rm gvs}}
\; &{\rm for}\;T_d\leq \tau_{\rm gvs}\\
\frac{\tau_{\rm gvs}}{2T_d}
\;&{\rm for}\;T_d> \tau_{\rm gvs}
\end{cases}
\qquad \qquad\text{for } \; g\to 0
\label{dNspont}
\eeqa
This function fits perfectly the curve 
obtained through numerical integration of Eq.(\ref{dNmeno2}) for small values of $g$ (open circles in Fig.\ref{fig_Nmeno}), and actually describes qualitatively  the  behaviour of $\frac{\braket{(\delta \hat N_-)^2}}{\braket{\hat N_s}+\braket{\hat N_i}}$  in all regimes, ranging from well below the MOPO threshold, to  close to it. In particular,  it shows that  the noise in the photon number difference is reduced by 50\% below shot-noise  at  $T_d =\tauGVS$, and then approaches asymptotically zero. 
\par 
In conclusion, the answer to our question is clear: in order to obtain sub-shot noise fluctuations in  the difference of the photon numbers a collection time  larger than   $\tauGVS$  is enough in any  regime. 
 This may appear perhaps surprising, because, according to the results presented in \cite{Corti2016},  close to threshold the cross correlation of twin beams 
  $ G^{(2)}_{si}(t,t')=\braket{\delta \hat  I_s(t)\delta \hat I_i(t') }     $  
acquires a slowly decaying  exponential tail 
$\approx e^{  -2g \epsilon \frac{ |t-t'|}{ \tauGVS} }$,  originating from  stimulated PDC in combination with backpropagation \cite{Corti2016}.
However, according to the results presented here, the temporal correlation of the intensity difference remains confined to the smaller time $\tauGVS$ characteristic of the spontaneous regime. 
Thus apparently  the stimulated processes do not contribute to the correlation of the intensity difference. Mathematically,this happens because  the same exponential  tails appear in the autocorrelation functions $G^{(2)}_{jj}(t,t')$ \cite{Corti2016} and cancel out the long-ranged part of the cross-correlation in the expression \eqref{Gmeno}. Intuitively,  in order to have a noise below the shot-noise of random processes, it is enough 
 to collect in the two arms all the twin photons originating from the same primary down-conversion processes, whose  delay  cannot exceed their transit time $\tauGVS$ across the slab.  
\section{Conclusions}
The analysis performed in this work has outlined several appealing features of the 
 countinuous variable entanglement of counter-propagating twin beams. Particularly appealing is the narrowband character of the emission( order  few  Ghz), the high-level of quadrature squeezing and the stability of the squeezing angle, which could make this monolithic source a viable alternative to the cavity OPO. 
\\
In contrast, the standard cavityless co-propagating configuration, as well known, is  highly multimode and broadband (order tens of  Thz).  While such a spectrally multimode entanglement may represent an important  resource for some applications, it is clear from our analysis that, especially in the high-gain,  it would be hard to detect squeezing inside the whole  bandwidth, because the squeezing-angle rotates rapidly with frequency, so that the contribution of the antisqueezed quadrature enters rapidly into play. Thus, contrary to what is usually thought(see e.g. \cite{Kaiser2016}), it would be probably hard to detect and exploit squeezing over the entire huge PDC bandwidth.  In contrast, the MOPO  offers a high stability of the squeezing angle, 
which can be seen  as a consequence of the  peculiar time scale $\tauGVS$ involved  in the counterpropagating interaction, which is much longer than the time scales $\tauGVM, \tauGVD$ characterizing co-propagating  light waves. 
\\
In the second part of the work we addressed the problem of the photon-number correlation in the MOPO. Perhaps surprisingly we have found that while the correlation time of twin beams has a critical divergence on approaching the threshold, the correlation of their intensity difference remains short ranged, and confined to the time $\tauGVS$ characteristic of the spontaneous regime. This result,  clearly positive for applications, means that sub-shot noise fluctuations of the photon-number difference can be measured  within a finite detection time. 
\appendix
\section{}
\label{Appendix}
 We derive here an  expression for the  noise of  the photon-number difference  $\hat N_-= \hat N_s -\hat N_i$, valid in  the limit $g\rightarrow 0$ (spontaneous regime). We use  the 
approximated expressions for the second-order field correlations  derived in Ref.\cite{Corti2016}  in the same limit 
\beqa
\braket{ \hat A_s^{\dagger\text{out}} (t_s)\hat A_s^{\text{out}} (t_s')}
&\approx&
\frac{g^2}{2\tau_{\rm gvs}}
{\rm Triangle}\left(\frac{t_s-t_s'}{2\tau_{\rm gvs}}\right)\label{Fcoh}\\
\braket{\hat A_s^{\text{out}} (t_s) \hat A^{\text{out}} _i(t_i)}
&\approx&
\frac{ge^{i\phi_p+ik_sl_c}}{2\tau_{\rm gvs}}
{\rm Rect}\left(\frac{t_s-t_i}{2\tau_{\rm gvs}}\right)
\label{Fpsi}
\eeqa
where:\\
- ${\rm Triangle}(x)=1-|x|$ if $|x|<1$,  ${\rm Triangle}(x)=0$ elsewhere is the triangular function, 
\\
- ${\rm Rect}(x)=1$ if $|x|<1/2$,  ${\rm Rect}(x)=0$ elsewhere is the rectangular box function, 
 \\
-we have omitted terms related to the slow phase $\beta(\Om)$ which are on the order $\tauGVM/\tauGVS$. \\
Accordingly,  the mean intensity  is  given by $\braket{\hat I_s}= \braket {\hat I_i} =\frac{g^2}{2\tau_{\rm gvs}}$. Then the shot-noise is $\braket {\hat N_s} + \braket {\hat N_i} = g^2\frac{  T_d}{\tau_{\rm gvs}}$. 
The intensity  correlation functions defined by Eqs.(\ref{Gjj}), (\ref{Gsi}) up to leading  order in $g \ll 1 $ become 
\beqa
G_{jj}(t,t')&=&\delta(t-t')\braket{I_j} + O(g^4)
\simeq \frac{g^2}{2\tau_{\rm gvs}}\delta(t-t') 
\label{Gjjbis}\\
G_{si}(t,t')
&\simeq& \frac{g^2}{4\tau_{\rm gvs}^2} {\rm Rect}\left(\frac{t-t'}{2\tau_{\rm gvs}}\right)
\label{Gsibis}
\eeqa
By inserting  the approximated expressions  \eqref{Gjjbis} and \eqref{Gsibis} into the correlation of the intensity difference in Eq.\eqref{Gmeno}, and evaluating the simple integrals involved in Eq.\eqref{dNmeno1}, we finally obtain
\beqa
\frac{\braket{(\delta \hat N_-)^2}}{\braket{\hat N_s}+\braket{\hat N_i}}
\simeq
\begin{cases}
1-\frac{T_d}{2\tau_{\rm gvs}}
\; &{\rm for}\;T_d\leq \tau_{\rm gvs}\\
\frac{\tau_{\rm gvs}}{2T_d}
\;&{\rm for}\;T_d> \tau_{\rm gvs}
\end{cases}
\qquad \qquad\text{for } \; g\to 0
\eeqa

\renewcommand\bibname{References}

\end{document}